\documentclass[aoas]{imsart}

\RequirePackage{amsthm,amsmath,amsfonts,amssymb}
\RequirePackage[authoryear]{natbib}
\RequirePackage[colorlinks,citecolor=blue,urlcolor=blue]{hyperref}
\RequirePackage{graphicx}
\usepackage{caption}
\usepackage{threeparttable}
\usepackage{subcaption}
\usepackage[ruled,vlined]{algorithm2e}

\startlocaldefs
\theoremstyle{plain}

\theoremstyle{remark}

\endlocaldefs

\begin{document}

\begin{frontmatter}
\title{BSNMani: Bayesian Scalar-on-network Regression with Manifold Learning}
\runtitle{BSNMani: Bayesian Scalar-on-network Regression with Manifold Learning}

\begin{aug}
\author[A]{\fnms{Yijun}~\snm{Li}\ead[label=e1]{liyijun@umich.edu}\orcid{0000-0003-0513-9565}},
\author[B]{\fnms{Ki Sueng}~\snm{Choi}\ead[label=e2]{kisueng.choi@mssm.edu}\orcid{0000-0002-6446-0677}},
\author[C]{\fnms{Boadie W.}~\snm{Dunlop}\ead[label=e3]{
bdunlop@emory.edu}\orcid{0000-0002-4653-0483}},
\author[C]{\fnms{W. Edward}~\snm{Craighead}\ead[label=e8]{
ecraigh@emory.edu}\orcid{0000-0001-9957-0027}},
\author[B]{\fnms{Helen S.}~\snm{Mayberg}\ead[label=e4]{
helen.mayberg@mssm.edu}\orcid{0000-0002-1672-2716}},
\author[E]{\fnms{Lana}~\snm{Garmire}\ead[label=e7]{lgarmire@med.umich.edu}\orcid{0000-0002-4654-2126}},
\author[D]{\fnms{Ying}~\snm{Guo}\ead[label=e5]{yguo2@emory.edu}\orcid{0000-0003-3934-3097}},
\author[A]{\fnms{Jian}~\snm{Kang}\ead[label=e6]{jiankang@umich.edu}\orcid{0000-0002-5643-2668}}
\address[A]{Department of Biostatistics,
University of Michigan\printead[presep={,\ }]{e1};\printead{e6}}
\address[B]{Icahn School of Medicine at Mount Sinai, New York, NY USA\printead[presep={,\ }]{e2};\printead{e4}}
\address[C]{Department of Psychiatry and Behavioral Sciences, Emory University School of Medicine\printead[presep={,\ }]{e3}\printead{e8}}
\address[E]{Department of Computational Medicine and Bioinformatics,
University of Michigan \printead[presep={,\ }]{e7}}
\address[D]{Department of Biostatistics and Bioinformatics,
Emory University\printead[presep={,\ }]{e5}}
\end{aug}

\begin{abstract}
Brain connectivity analysis is crucial for understanding brain structure and neurological function, shedding light on the mechanisms of mental illness. To study the association between individual brain connectivity networks and the clinical characteristics, we develop BSNMani: a Bayesian scalar-on-network regression model with manifold learning. BSNMani comprises two components: the network manifold learning model for brain connectivity networks, which extracts shared connectivity structures and subject-specific network features, and the joint predictive model for clinical outcomes, which studies the association between clinical phenotypes and subject-specific network features while adjusting for potential confounding covariates. For posterior computation, we develop a novel two-stage hybrid algorithm combining Metropolis-Adjusted Langevin Algorithm (MALA) and Gibbs sampling. Our method is not only able to extract meaningful subnetwork features that reveal shared connectivity patterns, but can also reveal their association with clinical phenotypes, further enabling clinical outcome prediction. We demonstrate our method through simulations and through its application to real resting-state fMRI data from a study focusing on Major Depressive Disorder (MDD). Our approach sheds light on the intricate interplay between brain connectivity and clinical features, offering insights that can contribute to our understanding of psychiatric and neurological disorders, as well as mental health.
\end{abstract}


\end{frontmatter}

\section{Introduction}
Network-valued data has become increasingly important in neuroimaging research in the past decade, especially in understanding brain structure, function, and its role in cognitive development, neuro-degenerative diseases, depression, and other mental conditions and illness (\cite{belmonte2004autism,supekar2008network,zhang2011disrupted,bullmore2009complex}). In the context of neuroimaging studies, network-valued data are typically derived from structural or functional imaging data. For example, functional brain connectivity networks can be generated from resting state functional magnetic resonance imaging (rs-fMRI) data, a common neuroimaging technology which measures small fluctuations in the blood oxygen level dependence in the brain when the subject is in resting state, not engaging in any particular task. Depending on the resolution of the specific technology platform, the rs-fMRI data can consist of between hundreds of thousands and millions of voxels. In neuroimaging analysis, these voxels are often categorized into well-defined and reproducible regions of interest (ROIs) based on established brain atlases (\cite{power2011functional,gordon2016generation,schaefer2018local}). Such ROIs have been shown to be associated with major brain functions, including motor control, reasoning, emotion regulation, etc. Functional connectivity can then be computed based on the pairwise correlation of time-series rs-fMRI data at the ROI level. Such connectivity data are very informative with respect to brain function and reveal insights into the underlying mechanism of neurological and psychiatric disorders.

There are several key challenges in analyzing brain connectivity data. Even after parcellation, an individual brain connectivity network can still contain hundreds of ROIs and therefore hundreds of thousands of edges. Various computational strategies have been developed to model such high-dimensional data. Some methods directly model the full set of edges in the brain connectivity network. For example, image-on-scalar regression models (\cite{zhu2014spatially,chen2016local,zhang2023image}) have image predictors and scalar responses and reveal the association between brain connectivity networks and clinical characteristics through spatially variable coefficients. Other works utilize network edges as covariates and clinical information as the outcome, attempting to identify edges or clusters of edges/nodes that are important for the clinical outcome. \cite{wang2019symmetric} proposed a symmetric bilinear regression model to identify small subgraphs in the network that are associated to the outcome. \cite{guha2021bayesian} proposed novel network shrinkage priors on the regression coefficients of network edge predictors in order to identify ROIs and interconnections that are significant for creativity measure. \cite{morris2022scalar} proposed a scalar-on-network model that leveraged known brain functional organization to identify edges that are useful for the clinical outcome. While capable of identifying node or edge clusters that are related the clinical outcome, such approaches cannot provide further information on the underlying functional structure of the clusters. Additionally, the high dimensionality of the full edge set requires some form of regularization, which can lead to potential loss of information.

Another strategy in network analysis treats the observed connectivity network as a summation of several underlying subnetworks. Such approaches often reduce the dimensionality of the network through dimension-reduction techniques. For example, \cite{sun2017store} and \cite{wang2017bayesian} treat brain connectivity networks as responses and clinical information as covariates, efficiently learning the regression coefficients through low-rank factorization. \cite{durante2017nonparametric} also employed low-rank factorization to model the population distribution of binary network-valued data. However, this model does not explicitly examine the association between network and clinical data. Furthermore, the thresholding procedure required to produce binary connectivity networks can lead to a greater loss of information. \cite{wang2019common,wang2023locus,amico2017mapping} model full brain connectivity networks by estimating them as a sum of either latent subspaces or latent connectivity traits. These methods focus on uncovering the latent structures that underlie brain connectivity, without explicitly modeling the associations between clinical outcome and connectivity. \cite{ma2022semi} adopts a two-stage scalar-on-network regression approach that first extracts lower-dimensional network features and subsequently studies the association between such features and clinical outcome. However, such sequential two-stage strategies can potentially cause label switching problems in the learned network features. 

\begin{figure}[h!]
    \centering
    \includegraphics[scale=0.3]{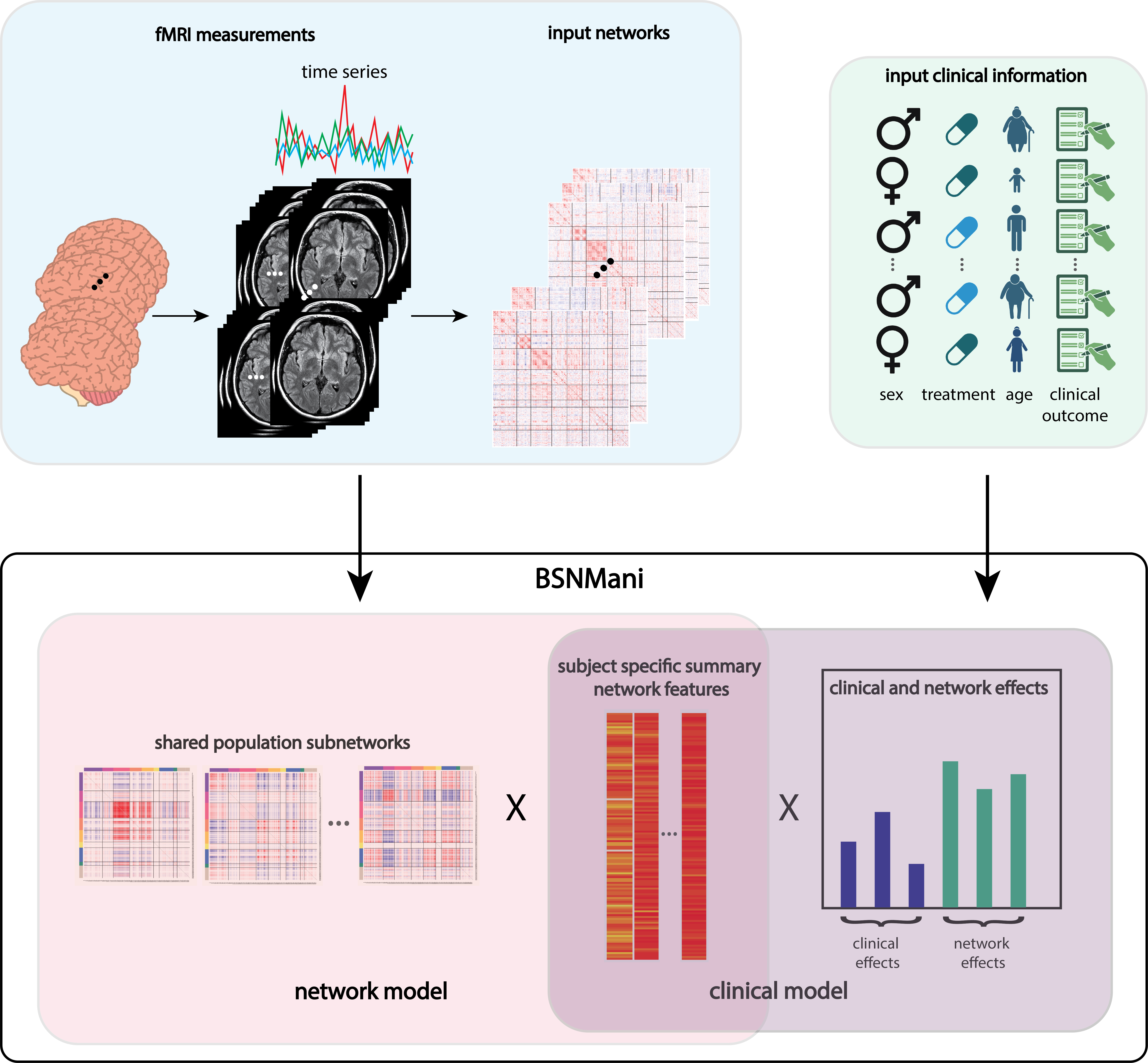}
    \caption{Workflow of BSNMani.}
    \label{fig:flow}
\end{figure}

Motivated by the aforementioned challenges, as well as the need for a fully data-driven, joint model framework to study the association between clinical data and brain connectivity, we propose BSNMani, a novel Bayesian scalar-on-network regression via manifold learning. BSNMani consists of two components: a network decomposition model where we learn both population-wide and subject-specific connectivity traits, and a clinical regression model where we study the association between a scalar clinical outcome and subject-specific network features while adjusting for potentially confounding clinical covariates (Fig. \ref{fig:flow}). We propose a novel joint posterior inference algorithm, allowing us to learn network features that are predictive of the clinical outcome of interest, hence identifying outcome-related network patterns. We demonstrate BSNMani's performance through a real data analysis of the resting state brain networks from the Predictors of Remission in Depression to Individual and Combined Treatments (PReDICT) study (\cite{dunlop2012predictors,dunlop2018differential}) for Major Depressive Disorder (MDD), and through thorough simulation studies based on synthetic and data-driven simulation. Compared to existing methods, BSNMani showed both higher predictive accuracy with regards to clinical outcome, as well as an intrinsic advantage in recovering meaningful underlying population-wide subnetworks that can elucidate clinical outcome related brain connectivity patterns sources.
\newpage
\section{Model}
\label{s:model}
Suppose the dataset consists of $M$ subjects. Let $i=1,2,\cdots,M$ be the subject index. Denote the functional brain connectivity data for subject $i$ as $\mathbf{Y}_i$. $\mathbf{Y}_i$ is a symmetric $N\times N$ matrix, where $N$ represents the number of regions of interests (ROIs). Each entry in the $j$-th row and the $k$-th column ($j,k=1,2,\cdots,N$) represents the connectivity strength between the region $j$ and the region $k$. For this project, the connectivity of a region to itself is not of interest, so the diagonal of each matrix $\mathbf{Y}_i$ is set to zero. Since $\mathbf{Y}_i$ is a symmetric matrix, we can capture all its information in either the upper or lower triangular portion. Here, we define the operation $\mathrm{vecl}(\cdot)$ that vectorizes the lower triangular portion of a matrix input. For example, given an arbitrary $n\times n$ square matrix $\mathbf{B}$, $\mathrm{vecl}(\mathbf{B})=[B_{2,1},\cdots,B_{n,1},B_{3,2}\cdots,B_{n,2},\cdots,B_{n,n-1}]^T$.  Besides connectivity data, suppose that we also observe clinical data for each subject. Denote the scalar clinical outcome as $C_i$. Furthermore, there may also be relevant clinical covariates, such as age, gender, etc., that are associated with the clinical outcome. We denote such covariates as $\mathbf{z}_i=[z_{i1},z_{i2},\cdots,z_{ir}]^T$, with $r$ representing the number of clinical covariates to adjust for. 

\subsection{Bayesian scalar-on-network regression with manifold learning}
We propose the following Bayesian scalar-on-network regression model with manifold learning to decompose brain connectivity matrices and jointly analyze the decomposed features with the clinical outcome. For subject $i=1,2,\cdots,M$, BSNMani consisted of the following two components: \begin{equation}\label{eqn_model_p1}
    \mathbf{Y}_i = \sum_{l=1}^q\lambda_{il}\mathbf{u}_l\mathbf{u}_l^T + \boldsymbol{\epsilon}_i
\end{equation}
\begin{equation}\label{eqn_model_p2}
    C_i = \sum_{l=1}^q\beta_{l}\lambda_{il} + \boldsymbol{\alpha}^T\mathbf{z}_i + \delta_i
\end{equation}

In the first component of the model (equation \ref{eqn_model_p1}), we decompose the functional connectivity matrix $\mathbf{Y}_i$ as a weighted sum of $q$ underlying subnetworks ($q<<N$). Specifically, each underlying subnetwork $l$ is represented as a rank-one matrix $\mathbf{u}_l\mathbf{u}_l^T$. Such subnetworks aim to capture underlying population-wide brain connectivity traits, revealing specific subsets of ROIs that tend to be correlated with respect to brain function. $\lambda_{il}$ is a scalar that represents a subject-specific summary feature for the subnetwork $l$, summarizing the contribution of the subnetwork $l$ to the observed connectivity of the subject $i$'s $\mathbf{Y}_i$. Therefore, in the context of our study's application, $\lambda_{il}$ summarizes the contribution of the population subnetwork $l$ to the observed brain connectivity of subject $i$. $\boldsymbol{\epsilon}_i$ is a symmetric matrix representing the random noise in $\mathbf{Y}_i$ not captured by the mean model. We assume that random noise follows a Gaussian distribution with variance parameter $\sigma^2$: $\mathrm{vecl}(\boldsymbol{\epsilon}_i) \sim \mathrm{N}(0,\sigma^2)$. The first component of the model can also be rewritten as
\begin{equation}
\mathbf{Y}_i=\mathbf{U}\mathbf{\Lambda}_i\mathbf{U}^T+\boldsymbol{\epsilon}_i
\end{equation}, where $\mathbf{U}=[\mathbf{u}_1,\cdots,\mathbf{u}_q]$ is an $N\times q$ matrix representing the set of $q$ generating vectors for the underlying population-wide connectivity subnetworks. Furthermore, $\mathbf{U}$ is also an orthogonal matrix ($\mathbf{U}^T\mathbf{U}=\mathbf{I}_q$). Such matrices have unique properties in that they belong to the Stiefel manifold. Specifically, the set of $N \times q$ matrices $\mathbf{A}$ such that $\mathbf{A}^T\mathbf{A}=\mathbf{I}_q$, where $\mathbf{I}_q$ is a $q\times q$ identity matrix, represents the Stiefel manifold denoted by $\mathcal{V}_{q,N}$. The Stiefel manifold $\mathcal{V}_{q,N}$ is the space of $q-$frames in the $N-$dimensional real Euclidean space $\mathbb{R}^N$. $\boldsymbol{\Lambda}_i$ is a diagonal matrix s.t. $\boldsymbol{\Lambda}_i=\mathrm{diag}(\boldsymbol{\lambda}_i)$, $\boldsymbol{\lambda}_i=[\lambda_{i1},\cdots,\lambda_{iq}]^T$ . $\boldsymbol{\Lambda}_i$ contains the subject specific brain connectivity network summary features. We denote the entire set of subject specific network features as $\{\boldsymbol{\lambda}_i\}_{i=1}^M$. Therefore, our network manifold model projects the observed connectivity matrices $\mathbf{Y}_i$'s onto the Stiefel manifold, using a set of basis established with $\{\mathbf{u}_l\mathbf{u}_l^T\}_{l=1}^q$, where the subnetwork connectivity score $u_{il}u_{jl}$ between nodes $i$ and $j$ directly measures the quantitative evidence of edges in the corresponding functional module pair. Furthermore, $\{\boldsymbol{\lambda}_i\}_{i=1}^M$'s represent the subjects' lower-dimensional coordinates after being projected onto the Stiefel manifold. 

In the second part of the model \eqref{eqn_model_p2}, we study the association between the subjects' lower-dimensional connectivity coordinates $\lambda_{il}$ and the clinical outcome $C_i$ while adjusting for additional clinical covariates $\mathbf{z}_i$ in a linear regression model. $\boldsymbol{\beta}=[\beta_1,\cdots,\beta_q]$ is the regression coefficient that captures the effect of the subject network features. Besides connectivity network features, certain additional clinical covariates could also affect the clinical outcome in question. Therefore, we further adjust for the supplementary clinical covariates $\mathbf{z}_i$, whose effect is captured by the regression coefficient $\boldsymbol{\alpha}$. The term $\delta_i$ represents random noise not captured by the clinical model. Similarly to the network decomposition model, we also assume that the random noise in the clinical model follows a Gaussian distribution with variance parameter $\tau^2$: $\delta_i \sim \mathrm{N}(0,\tau^2)$.

The combined two components described above make up our overall scalar-on-network regression model. To avoid potential label-switching with respect to latent subnetworks, a frequent side effect of two-stage models, we propose a Bayesian joint modeling approach, where the model parameters are inferred not just based on observed connectivity data, but also based on clinical data. Furthermore, such an approach gives us deeper insight into how connectivity subnetworks associate with clinical phenotypes. 

\subsection{Posterior Inference} 
\begin{figure}[h!]
\centering
\includegraphics[scale=0.4]{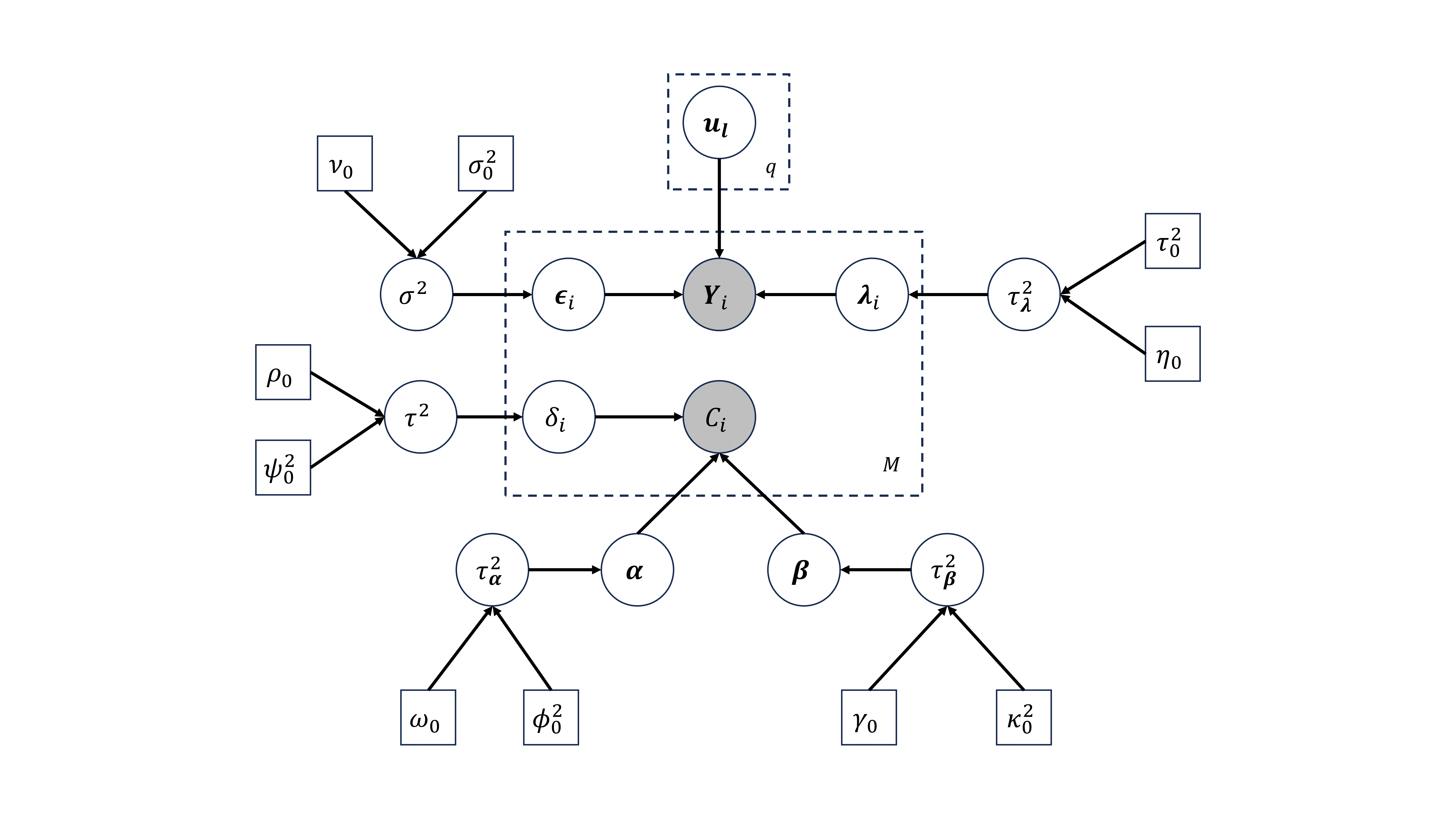}
\caption{Diagram of the BSNMani model. \textit{Note: gray circle represents observed data; white circle represents model parameters; white squares represent hyperparameters}.}
\label{diagram}
\end{figure}

In Figure \ref{diagram} we provide a diagram of BSNMani. For the prior of the orthogonal matrix parameter $\mathbf{U}$, we assume that it is uniformly distributed on the Stiefel manifold $\mathcal{V}_{q,N}$. For subject-specific random effects $\lambda_{il}$, we assign a conjugate normal prior $\lambda_{il}\sim \mathrm{N}(0,\tau^2_{\boldsymbol{\lambda}})$ and inverse gamma prior for its variance parameter $\tau^2_{\lambda}$. We assign conjugate normal priors for the regression coefficients $\boldsymbol{\beta}$ and $\boldsymbol{\alpha}$, and inverse gamma priors for their corresponding prior variance parameters $\tau^2_{\beta}$ and $\tau^2_{\alpha}$. Finally, since both the functional connectivity network model (part one) and the clinical outcome prediction model (part two) have normal random noise, we assigned conjugate inverse-gamma priors for their respective variance terms $\sigma^2$ and $\tau^2$. 
\begin{align*}
    &\mathbf{U} \sim \mathrm{Uniform}(\mathcal{V}_{q,N}); &&\boldsymbol{\lambda}_i \sim \mathrm{MVN}(\mathbf{0}, \tau^2_{\boldsymbol{\lambda}}\mathbf{I}_q); \\
    & 1/\tau^2_{\boldsymbol{\lambda}} \sim \mathrm{Gamma}(\eta_0/2, \eta_0\tau^2_0/2); && \boldsymbol{\beta} \sim \mathrm{MVN}(\mathbf{0}, \tau^2_{\beta}\mathbf{I}_q); \\
    &1/\tau^2_{\boldsymbol{\beta}} \sim \mathrm{Gamma}(\gamma_0/2, \gamma_0\kappa_0^2/2); && \boldsymbol{\alpha} \sim \mathrm{MVN}(\mathbf{0}, \tau^2_{\alpha}\mathbf{I}_T);\\
    &1/\tau^2_{\boldsymbol{\alpha}} \sim \mathrm{Gamma}(\omega_0/2, \omega_0\phi_0^2/2); &&1/\tau^2 \sim \mathrm{Gamma}(\rho_0/2,\rho_0\psi_0^2/2); \\
    & 1/\sigma^2 \sim \mathrm{Gamma}(\nu_0/2, \nu_0\sigma^2_0/2)
\end{align*}

For posterior inference of the model parameters, we consider a hybrid MALA-Gibbs algorithm (Algorithm \ref{alg:MALA-Gibbs}). To obtain MCMC samples of the model parameters, we adopt a parallel update strategy rather than a two-stage estimation strategy. Such an updating scheme allows for the borrowing of information between models when estimating the model parameters. 

Most of the parameters in BSNMani have closed-form posterior distributions, which can be sampled via Gibbs sampling. However, we do not have a closed-form posterior distribution for $\mathbf{U}$. The target posterior density of $\mathbf{U}$ is:

\begin{equation}
    \mathrm{\log}\pi(\mathbf{U}|\sim) \propto \mathrm{\log}\pi(\mathbf{U})+\sum_{i=1}^Mlog\pi(\mathbf{Y}_i|\sim) \propto \mathrm{trace}[(\sigma^{-2}\sum_{i=1}^M(\mathbf{U}\boldsymbol{\Lambda}_i\mathbf{U}^T\mathbf{Y}_i)].
\end{equation} Given the orthogonality restriction on $\mathbf{U}$, it can be difficult to directly obtain MCMC samples from the Stiefel manifold using standard MCMC procedures. Adapting the work of \cite{jauch2021monte}, we avoid such restrictions by constructing a target density on the unconstrained Euclidean space. \cite{jauch2021monte} introduced a bijection between the orthogonal matrix $\mathbf{U}$ and a corresponding full-rank matrix $\mathbf{X}$ of the same dimensions on the Euclidean space. Suppose $\mathbf{X}$ is a full-rank matrix such that $\mathbf{X} \in \mathbb{R}^{N,q}$, there exists a unique polar decomposition for $\mathbf{X}$ as \begin{equation}
    \mathbf{X} = \mathbf{U}_{\mathbf{X}}\mathbf{S}^{\frac{1}{2}}_{\mathbf{X}}
\end{equation}
where $\mathbf{U}_{\mathbf{X}} = \mathbf{X}(\mathbf{X}^T\mathbf{X})^{-\frac{1}{2}}$ and $\mathbf{S}_{\mathbf{X}}=\mathbf{X}^T\mathbf{X}$. It is therefore straightforward that $\mathbf{U}_{\mathbf{X}}$ is an orthogonal matrix. Furthermore, the matrix $\mathbf{X}$ follows a matrix normal distribution $\mathbf{X} \sim \mathrm{N}_{N,q}(\mathbf{0};\mathbf{I}_N, \mathbf{I}_q)$, if and only if the corresponding $\mathbf{U}_{\mathbf{X}}$ and $\mathbf{S}_{\mathbf{X}}$ are independent, $\mathbf{U}_{\mathbf{X}}\sim\mathrm{Uniform}(\mathcal{V}_{q,N})$, and $\mathbf{S}_{\mathbf{X}}\sim \mathrm{Wishart}_q(N,\boldsymbol{\Sigma})$. Therefore, the matrix parameter $\mathbf{U}$ in our scalar-on-network regression model, to which we assign a uniform prior on the Stiefel manifold $\mathcal{V}_{q,N}$, corresponds to an $N\times q$ full rank matrix $\mathbf{X}$ on the unconstrained Euclidean space that follows a matrix variate normal distribution: $\mathbf{X} \sim \mathbf{N}_{N,q}(\mathbf{0};\mathbf{I}_N,\mathbf{I}_q)$. 

Thus, we can construct a transformed posterior target density for $\mathbf{X}$: \begin{equation}\label{X_pos}
    \mathrm{\log}\pi(\mathbf{X}|\sim) \propto -\frac{1}{2}\mathrm{trace}(\mathbf{X}^T\mathbf{X})+\frac{1}{\sigma^2}\cdot\mathrm{trace}\left(\sum_{i=1}^M\boldsymbol{\Lambda}_i\mathbf{U}_{\mathbf{X}}\mathbf{Y}_i\mathbf{U}_{\mathbf{X}}\right).
\end{equation}
 Since $\mathbf{X}$ is an unconstrained matrix, it is compatible with standard MCMC sampling algorithms. 
First, we can directly sample $\{\mathbf{X}^{(1)},\mathbf{X}^{(2)},\cdots,\mathbf{X}^{(K)}\}$ from the constructed target density of $\mathbf{X}$ (Equation \ref{X_pos}) using the Metropolis-adjusted Langevin Algorithm (MALA), then obtain approximate MCMC samples of $\mathbf{U}$: $\{\mathbf{U}_{\mathbf{X}}^{(1)},\mathbf{U}_{\mathbf{X}}^{(2)},\cdots,\mathbf{U}_{\mathbf{X}}^{(K)}\}$ via polar expansion.
\begin{algorithm}[hbt!]
\caption{BSNMani algorithm}\label{alg:MALA-Gibbs}
\SetAlgoLined
\KwData{$\{\mathbf{Y}_i,C_i,\mathbf{z}_i\}_{i=1}^M$; hyperprior parameters: $\rho_0$, $\psi^2_0$, $\nu_0$, $\sigma^2_0$, $\eta_0$, $\tau^2_0$, $\gamma_0$, $\kappa^2_0$, $\omega_0$, $\phi_0^2$; number of iterations $K$; target acceptance rate $\rho$; number of iterations between tuning: $K_0$}
\KwResult{posterior samples of $\mathbf{X}$, $\mathbf{U}$, $\{\boldsymbol{\lambda}_i\}_{i=1}^M$, $\sigma^2$, $\boldsymbol{\beta}$, $\boldsymbol{\alpha}$, $\tau^2$, $\tau^2_{\boldsymbol{\lambda}}$, $\tau^2_{\boldsymbol{\beta}}$, $\tau^2_{\boldsymbol{\alpha}}$}
\SetKwBlock{Begin}{Begin}{}
\Begin{
Initialize $\mathbf{X}^{(0)}$, $\mathbf{U}^{(0)}$, $\{\boldsymbol{\lambda}^{(0)}_i\}_{i=1}^M$, $\sigma^{2(0)}$, $\boldsymbol{\beta}^{(0)}$, $\boldsymbol{\alpha}^{(0)}$, $\tau^{2(0)}$, $\tau^{2(0)}_{\boldsymbol{\lambda}}$, $\tau^{2(0)}_{\boldsymbol{\beta}}$, $\tau^{2(0)}_{\boldsymbol{\alpha}}$, and initial stepsize for MALA: $\omega^{(0)}$\;
\SetKwFor{For}{for}{do}{}
\For{$k=1,\cdots,K$}{
  \SetKwFor{For}{for}{do}{end}
  Update $\boldsymbol{\lambda}_i^{(k)}$, $\mathbf{d}^{(k)}$, $\sigma^{2(k)}$, $\tau^{2(k)}$, $\tau^{2(k)}_{\boldsymbol{\lambda}}$, $\tau^{2(k)}_{\boldsymbol{\beta}}$, $\tau^{2(k)}_{\boldsymbol{\alpha}}$ from their corresponding conjugate posterior distributions via Gibbs sampling;\\
  Update $\mathbf{X}^{(k)}$ via MALA \begin{itemize}
        \item Generate $\mathbf{D} \in \mathbb{R}^{N,q} \overset{\mathrm{iid}}{\sim} \mathrm{N}(0,1)$ and $V\sim \mathrm{Uniform}(0,1)$.
        \item Set $\mathbf{Q} = \mathbf{X}^{(k-1)}+\frac{\omega^2}{2}\nabla \mathrm{\log}\pi(\mathbf{X}^{(k-1)})+\omega\mathbf{D}$
        \item $\mathbf{X}^{(k)}=\begin{cases}
    &\mathbf{Q}; \\
    &\quad \text{if } V<\mathrm{\min}\{1,\frac{\pi(\mathbf{Q})}{\pi(\mathbf{X}^{(k-1)})}\frac{\exp(-||\mathbf{X}^{(k-1)}-\mathbf{Q}-\omega^2\nabla \mathrm{\log}\pi(\mathbf{Q}|\sim)/2||^2/2\omega^2)}{\exp(-||\mathbf{Q}-\mathbf{X}^{(k-1)}-\omega^2\nabla \mathrm{\log}\pi(\mathbf{X}^{(k-1)}|\sim)/2||^2/2\omega^2)}\}\\
    &\mathbf{X}^{(k-1)};\\
    &\quad \text{otw}
    \end{cases}$
    \end{itemize}     
    Update $\mathbf{U}^{(k)} = \mathbf{X}^{(k)}(\mathbf{X}^{T(k)}\mathbf{X}^{(k)})^{-\frac{1}{2}}$\;
    \If{$k$ mod $K_0==0$}{
      Compute empirical acceptance rate $\rho'$\;
      \eIf{$\rho'<\rho$}{$\omega^{(k/K_0)}=\omega^{(k/K_0-1)}\cdot 0.9$}{$\omega^{(k/K_0)}=\omega^{(k/K_0-1)}\cdot 1.1$}
    }   
    }
    }
\end{algorithm}

In real data analysis of functional brain connectivity data, the observed connectivity matrices can include 200 to 500 ROIs depending on the reference brain atlas used. Therefore, as the number of ROIs rises, convergence for the joint posterior sampling scheme proposed above can become slow as the algorithm tries to balance the data likelihood from both the network and the clinical model. To speed up MCMC convergence, we propose the following approximate two-stage posterior sampling algorithm. 

Denote the full set of model parameters as $\boldsymbol{\Theta}$,  network-model-specific parameters as $\boldsymbol{\Theta}_{\mathbf{Y}}$, and clinical-model-specific parameters as $\boldsymbol{\Theta}_{C}$ such that \begin{align*}
    &\boldsymbol{\Theta}=\{\boldsymbol{\Theta}_{Y},\{\boldsymbol{\lambda}_i\}_{i=1}^M,\boldsymbol{\Theta}_{C},\tau^2_{\boldsymbol{\lambda}},\tau^2_{\boldsymbol{\alpha}},\tau^2_{\boldsymbol{\beta}}\};\quad \boldsymbol{\Theta}_{Y} = \{\mathbf{U},\sigma^2\}; \quad \boldsymbol{\Theta}_{C} = \{\boldsymbol{\alpha},\boldsymbol{\beta},\tau^2\}.
\end{align*}
The joint target posterior density can then be written as a product between the target posterior density of the network decomposition model and the clinical regression model: $$f(\boldsymbol{\Theta}) \propto \left[\prod_{i=1}^M\pi(\mathbf{Y}_i|\boldsymbol{\Theta}_{Y},\boldsymbol{\lambda}_i)\pi(C_i|\boldsymbol{\Theta}_{C},\boldsymbol{\lambda}_i)\pi(\boldsymbol{\lambda}_i|\tau^2_{\boldsymbol{\lambda}})\right]\pi(\tau^2_{\boldsymbol{\lambda}})\pi(\boldsymbol{\Theta}_{Y})\pi(\boldsymbol{\Theta}_{C}|\tau^2_{\boldsymbol{\alpha}},\tau^2_{\boldsymbol{\beta}})\pi(\tau^2_{\boldsymbol{\alpha}})\pi(\tau^2_{\boldsymbol{\beta}}).$$
To speed up model convergence, we propose a proposal density $g(\boldsymbol{\Theta})$, which is a product of the posterior density for the network model denoted as $g_1(\boldsymbol{\Theta_{\boldsymbol{Y}}})$, and the marginal clinical model posterior density $g_2(\boldsymbol{\Theta_{C}})$, where $\{\boldsymbol{\lambda}_i\}_{i=1}^M$ are treated as fixed input data instead of variables: \begin{align*}
  g(\boldsymbol{\Theta}) &\propto g_1(\boldsymbol{\Theta}_{\mathbf{Y}})\cdot g_2(\boldsymbol{\Theta}_C)\  \mbox{with}\  \\
  g_1(\boldsymbol{\Theta}_{\mathbf{Y}}) & \propto  \left[\prod_{i=1}^M\pi(\mathbf{Y}_i|\boldsymbol{\Theta}_{Y},\boldsymbol{\lambda}_i)\pi(\boldsymbol{\lambda}_i|\tau^2_{\boldsymbol{\lambda}})\right]\pi(\tau^2_{\boldsymbol{\lambda}})\pi(\boldsymbol{\Theta}_{Y})\ \mbox{and}\ \\
  g_2(\boldsymbol{\Theta}_C) & \propto \left[\prod_{i=1}^M\pi(C_i|\boldsymbol{\Theta}_{C},\boldsymbol{\lambda}_i)\right]\pi(\boldsymbol{\Theta}_{C}|\tau^2_{\boldsymbol{\alpha}},\tau^2_{\boldsymbol{\beta}})\pi(\tau^2_{\boldsymbol{\alpha}})\pi(\tau^2_{\boldsymbol{\beta}})*\frac{1}{A(\boldsymbol{\lambda})}. 
\end{align*}
Such a strategy allows for faster convergence by allowing the random effects parameters $\{\boldsymbol{\lambda}_i\}_{i=1}^M$ to be updated using only the likelihood of the network data. 

When sampling from $g_1(\boldsymbol{\Theta}_{\mathbf{Y}})$, we further decompose $g_1(\boldsymbol{\Theta}_{\mathbf{Y}})$ into a product of $g_{1,1}(\boldsymbol{\Theta}_{\mathbf{Y}})$ and $g_{1,2}(\boldsymbol{\Theta}_{\mathbf{Y}})$:  \begin{align*}
    g_1(\boldsymbol{\Theta}_{\mathbf{Y}}) &= \left[\prod_{i=1}^M\pi(\mathbf{Y}_i|\boldsymbol{\Theta}_{Y},\boldsymbol{\lambda}_i)\pi(\boldsymbol{\lambda}_i|\tau^2_{\boldsymbol{\lambda}})\right]\pi(\tau^2_{\boldsymbol{\lambda}})\pi(\boldsymbol{\Theta}_{Y})\\
    &= \underline{\left[\prod_{i=1}^M\pi(\mathbf{Y}_i|\boldsymbol{\Theta}_{Y})\right]\pi(\boldsymbol{\Theta}_{Y})}\cdot\underline{\left[\prod_{i=1}^M\pi(\boldsymbol{\lambda}_i|\tau^2_{\boldsymbol{\lambda}},\mathbf{Y}_i,\boldsymbol{\Theta}_Y)\right]\pi(\tau^2_{\boldsymbol{\lambda}}|\boldsymbol{\lambda}_1,\cdots,\boldsymbol{\lambda}_M)},\\
    &\qquad g_{1,1}(\boldsymbol{\Theta}_{\mathbf{Y}}) \qquad \qquad \qquad \qquad g_{1,2}(\boldsymbol{\Theta}_{\mathbf{Y}})
\end{align*}
 where $\prod_{i=1}^M\pi(\mathbf{Y}_i|\boldsymbol{\Theta}_{Y}) = \int\left[\prod_{i=1}^M\pi(\mathbf{Y}_i|\boldsymbol{\Theta}_Y,\boldsymbol{\lambda}_i)\pi(\boldsymbol{\lambda}_i|\tau^2_{\boldsymbol{\lambda}})d\boldsymbol{\lambda}_i\right]\pi(\tau^2_{\boldsymbol{\lambda}})d\tau^2_{\boldsymbol{\lambda}}$.  The term $g_{1,1}(\boldsymbol{\Theta}_{\mathbf{Y}})$ integrates out the random effects variables $\{\boldsymbol{\lambda}_i\}_{i=1}^M$ and their variance parameter $\tau^2_{\boldsymbol{\lambda}}$. This allows us to obtain posterior samples of  $\boldsymbol{\Theta}_{\mathbf{Y}}$ from  $g_{1,1}(\boldsymbol{\Theta}_{\mathbf{Y}})$, and $\{\boldsymbol{\lambda}_i\}_{i=1}^M$ and $\tau^2_{\boldsymbol{\lambda}}$ from $g_{1,2}(\boldsymbol{\Theta}_{\mathbf{Y}})$ in parallel, further speeding up convergence.

In $g_2(\boldsymbol{\Theta_{C}})$, the subject-specific random effects $\{\boldsymbol{\lambda}_i\}_{i=1}^{M}$ are treated as fixed input data rather than as variables. Therefore, the normalizing constant for $g_2(\boldsymbol{\Theta}_C)$ needs to be updated to reflect such change.  We denote the new normalizing constant of the $g_2(\boldsymbol{\Theta}_C)$ as  $A(\boldsymbol{\lambda})$ . Furthermore, the ratio between the target density and our new proposal density is:\begin{align*}
    \frac{f(\boldsymbol{\Theta})}{g(\boldsymbol{\Theta})} = A(\boldsymbol{\lambda})= \int\left[\prod_{i=1}^M\pi(C_i|\boldsymbol{\Theta}_{C},\boldsymbol{\lambda}_i)\right]\pi(\boldsymbol{\Theta}_{C}|\tau^2_{\boldsymbol{\alpha}},\tau^2_{\boldsymbol{\beta}})\pi(\tau^2_{\boldsymbol{\alpha}})\pi(\tau^2_{\boldsymbol{\beta}})d\boldsymbol{\Theta}_{C}d\tau^2_{\boldsymbol{\alpha}}d\tau^2_{\boldsymbol{\beta}}.
\end{align*}

Therefore, we can induce a two-stage sampling strategy to obtain approximate joint modeling samples via the Metropolis Hastings algorithm. First, we obtain posterior samples of $\boldsymbol{\Theta}_{\mathbf{Y}}$, $\{\boldsymbol{\lambda}_i\}_{i=1}^M$, and $\tau^2_{\boldsymbol{\lambda}}$ from $g_1(\boldsymbol{\Theta}_{\mathbf{Y}})$ in parallel. Next, we jointly sample $\boldsymbol{\Theta}_{C},\tau^2_{\boldsymbol{\alpha}},\tau^2_{\boldsymbol{\beta}}$ from $g_2(\boldsymbol{\Theta}_C)$, conditioning on the samples of $\{\boldsymbol{\lambda}_i\}_{i=1}^M$ obtained from $g_1(\boldsymbol{\Theta}_{\mathbf{Y}})$. Finally, we obtain approximate samples from the target joint posterior density $f(\boldsymbol{\Theta})$ via the independent Metropolis Hastings algorithm, using $A(\boldsymbol{\lambda})$ to compute the acceptance probability. 

\section{Simulations}
\begin{figure}[h!]
\centering
\includegraphics[scale=0.3]{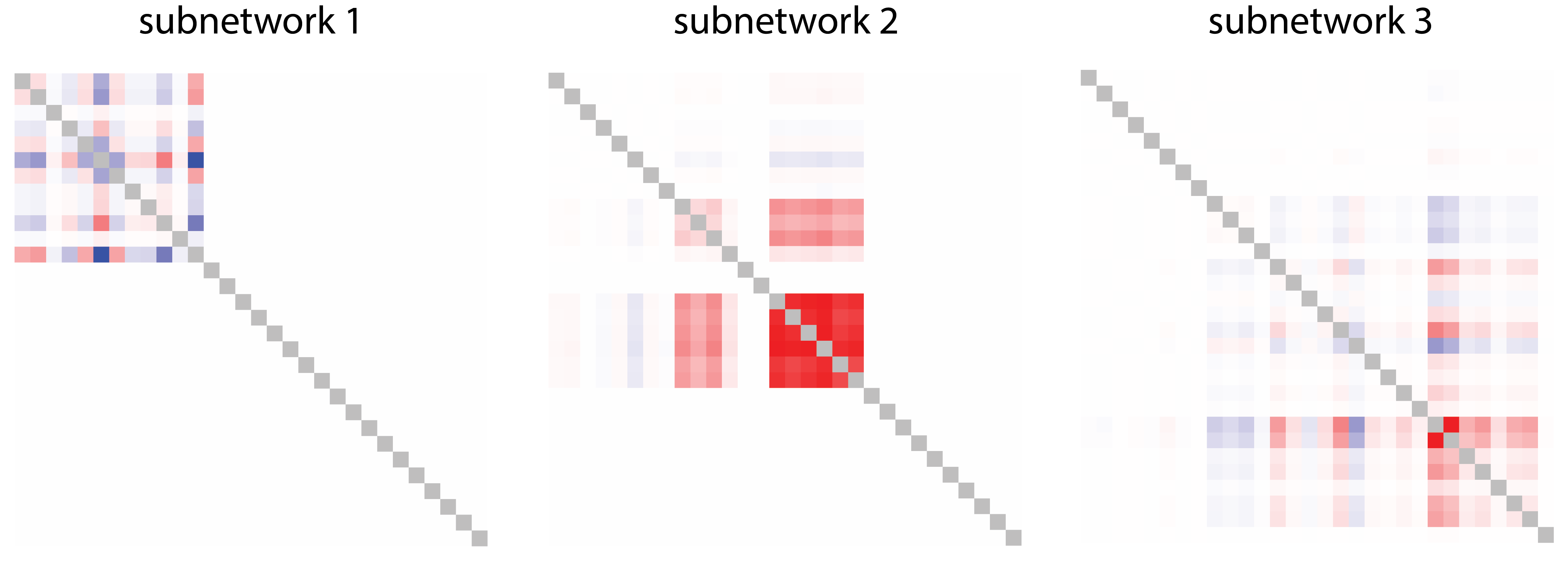}
\caption{Ground truth underlying true subnetworks for the small scale synthetic data simulation.}
\label{sim}
\end{figure}

In this section, we demonstrate the parameter estimation performance and the predictive performance of BSNMani through simulation studies. We compare the performance of our model with existing methods. We consider full edge set regression approaches such as GroupBoosting (\cite{morris2022scalar}), Support Vector Regression (SVR), lasso, and elastic net. Furthermore, we consider other subnetwork methods that first extract latent connectivity traits/subnetworks and then predict the clinical outcomes based on subjects' loadings on these subnetworks. Specifically, we adopts the recently developed LOCUS method \cite{wang2023locus}, which is a sparse blind source separation method that extracts latent connectivity traits from observed brain connectivity data without utilizing the clinical information. Given that LOCUS is not designed for prediction purposes, we incorporate a subsequent linear regression (LR) step to forecast the clinical outcome using subjects' trait loadings derived from LOCUS.  The method was implemented using the "LOCUS" R package on CRAN. We tried implementing LOCUS with no sparsity regularization, as well as LOCUS with the SCAD and L1 regularization. First, we investigated the robustness of BSNMani under varying population sizes and varying levels of signal-to-noise ratios via a small-scale simulation study consisting of synthetic datasets. Our synthetic data simulation procedure assumes that the hypothetical patient cohort shares a common latent connectivity structure ($\mathbf{U}$), with individual-specific summary subnetwork features ($\{\boldsymbol{\lambda}_i\}_{i=1}^M$). We set the number of ROIs to 30 ($N=30$) and the number of true latent subnetworks to 3 ($q=3$). For this study, we simulated ground truth latent subnetworks with simple block diagonal structures (see Figure \ref{sim}). We randomly generated subject-specific subnetwork summary features $\{\boldsymbol{\lambda}_i\}_{i=1}^M$ from the exponential distribution and generated synthetic network data based on our network model specification. Similarly, we generated a synthetic continuous covariate and a binary covariate as synthetic clinical covariates. We then set the regression coefficients at arbitrary ground truth values and then generated synthetic clinical outcomes based on our clinical regression model. To evaluate the robustness of BSNMani, we simulated data with different levels of the overall signal-to-noise ratio (SNR) in the network (denoted as $SNR_Y$) and clinical data (denoted as $SNR_C$) at 1 and 3. We define SNR as the ratio between the variance of the values in the mean model ($\mathbf{U}\boldsymbol{\Lambda}_i\mathbf{U}^T$ for the network model and $\boldsymbol{\beta}^T\boldsymbol{\lambda}_i+\boldsymbol{\alpha}^T\mathbf{z}_i$ for the clinical model) and the variance in the random noise ($\boldsymbol{\epsilon}_i$ for the network model and $\delta_i$ for the clinical model). We also varied the overall population size $M$ to 390, 650, and 1300. The entire simulated dataset was split into training and testing sets with the corresponding sizes 300, 500, 1000 and 90, 150, 300. The overall performance of the model was evaluated based on the model parameter estimation error and the comparison of the prediction error with existing methods. Table \ref{tab:par-estimation-1} shows that BSNMani maintains relatively low parameter estimation root mean square error (RMSE) under different population sizes and SNR in network and clinical data. Specifically, as the sample size and network SNR increase, the accuracy of the estimation of the subject-specific subnetwork features parameter $\{\boldsymbol{\lambda}_i\}_{i=1}^M$ and the network residual variance parameter $\sigma^2$ increase. The estimation accuracy of the matrix parameter $\mathbf{U}$, the regression coefficients $\mathbf{d}=[\boldsymbol{\beta}^T,\boldsymbol{\alpha}^T]^T$ and clinical data residual variance parameter $\tau^2$ also increase as the level of SNR in network data and clinical data increases, respectively. BSNMani also maintained high predictive accuracy under varying simulation settings, with the predictive $R^2$ generally increasing as the population size and SNR in clinical data increase (see Table \ref{tab:model-based-sim}). Furthermore, BSNMani demonstrated advantages with regard to predictive accuracy compared to all existing methods that were implemented in the comparison. This advantage becomes increasingly distinctive as the SNR levels in network data, clinical data, and population size increase. 

\begin{table}[h!]
\caption{Predictive $R^2$ of BSNMani, SVR, lasso, elastic net, GroupBoosting, and LOCUS+LR* under default SCAD regularization, L1 regularization, and no regularization, respectively, in the small-scale synthetic simulation results under varying population size, network and clinical signal-to-noise ratios.} 
\resizebox{5.5in}{!}{
    \begin{tabular}{lll|llllllll}
    \hline \hline
    \multicolumn{3}{c}{Simulation setting} & \multicolumn{8}{c}{Predictve $R^2$} \\ \cline{1-3} \cline{4-11}
    $M$ & $SNR_Y$ & $SNR_C$  & BSNMani & SVR & lasso & elastic net & GroupBoosting & LOCUS (SCAD) + LR & LOCUS (L1) + LR & LOCUS + LR\\    \hline
    300 & 0.5 & 3 & 0.5387 & 0.3032 & 0.4826 & 0.4890 & 0.4529 & 0.5216 & 0.5215 &0.5227  \\ \hline
    500 & 0.5 & 3 & 0.4846 & 0.2894 & 0.4727 & 0.4754 & 0.4174 & 0.4766 & 0.4764&  0.4773  \\ \hline 
    1000 & 0.5 &3 & 0.6170 & 0.4644 & 0.5943 & 0.5977 &0.4992 &0.6134 &0.6131 & 0.6132 \\ \hline 
    500 & 0.5 & 6 & 0.8120 & 0.5808 &0.7981 & 0.7988 & 0.7229 & 0.8017 & 0.8017 & 0.8024  \\ \hline 
    500 & 0.1 & 6 & 0.8228 & 0.4150 & 0.7909 &0.7909 & 0.7085 & 0.7918 & 0.7907 & 0.7919  \\
    \hline \hline
    \end{tabular}
}
\begin{tablenotes}
      \small
      \item LOCUS+LR: a two-step composite pipeline that first uses LOCUS to extract latent connectivity traits from observed brain connectivity data using without utilizing the clinical information, then predicts the clinical outcome using subjects' trait loadings derived from LOCUS using linear regression.
    \end{tablenotes}
\label{tab:model-based-sim}
\end{table}
\newpage
Our data-driven simulation utilizes the learned subnetworks and subject-specific subnetwork features from the PReDICT study to generate realistic brain connectivity network and clinical data. To generate realistic clinical data, we estimate the SNR in real clinical data in the PReDICT study and adjusted the random noise variance $\tau^2$ so that the SNR level in the simulated clinical data is at a similar level. Specifically, the SNR in the simulated clinical data is set at $SNR_C=0.38$. Similarly, we also computed the region-pair-wise SNR in real brain connectivity networks in the PReDICT study. To account for heterogeneity in random noise levels at different region pairs, we simulated random noise by setting region-pair specific noise variance rather than uniform variance, as BSNMani assumes. We vary the mean level of region-pair-specific SNR to further examine BSNMani's robustness under increased noise levels. Specifically, we simulate connectivity networks based on mean network signal-to-noise ratios at 3, 1, 0.5, and 0.0416 (around that of the real brain network data). As with the small-scale synthetic data simulation, we split our data into training and testing sets and evaluated BSNMani performance based on parameter estimation accuracy in the training set, and compared BSNMani's predictive performance with existing methods. Due to the larger number of ROIs, we excluded SVR and GroupBoosting from the comparison. Table \ref{tab:par-estimation-2} shows that BSNMani is able to accurately recover ground truth parameter values, even when the data is noisy and mis-specified from the model. BSNMani recovered the network parameters very well at different levels of network signal-to-noise ratios, including at mean SNR levels around that of the observed network data. The estimation accuracy of the clinical regression coefficient $\mathbf{d}$ and the clinical outcome variance parameter $\tau^2$ decreased as the clinical signal-to-noise ratio decreased. Furthermore, the predictive performance of BSNMani also showed an advantage over that of existing methods at different levels of mean $SNR_Y$ and $SNR_C$ (see Table \ref{tab:data-driven-sim}). Specifically, BSNMani's predictive $R^2$ was distinctively higher than those of full edge regression methods, namely lasso and elastic net; such advantage becomes more striking as the signal of noise ratio decreases in the simulated data. Moreover, BSNMani also achieved better than the composite LOCUS+LR approach.
\begin{table}[h!] 
\caption{Predictive $R^2$ of BSNMani, lasos, elastic net, and LOCUS+LR* under default SCAD regularization, L1 regularization, and no regularization, respectively, in the data-driven simulation results under varying mean network and clinical signal-to-noise ratios.} 
\resizebox{5.5in}{!}{
    \begin{tabular}{ll|llllll}
    \hline \hline
    \multicolumn{2}{c}{Simulation setting} &\multicolumn{6}{c}{Predictve $R^2$} \\ \cline{1-2} \cline{3-8}
    $SNR_Y$ & $SNR_C$ & BSNMani &  lasso & elastic net &  LOCUS (SCAD) + LR & LOCUS (L1) + LR & LOCUS + LR\\    \hline
    0.0461 & 3 & 0.703 & 0.617& 0.596& 0.686& 0.686& 0.686   \\ \hline
    0.5 & 3 & 0.702 & 0.645&0.644 &0.699 &0.699 & 0.699    \\ \hline 
    1 & 3 & 0.702& 0.653&0.655 &0.700 & 0.700&0.700   \\ \hline 
    3 & 3 & 0.7022 & 0.614 & 0.623& 0.700&0.700 &0.700    \\ \hline 
    3 & 0.38 & 0.111 & 0.040& 0.049& 0.009&0.009 &0.009    \\ \hline 
    3 & 0.5 & 0.159 & 0.077& 0.079& 0.145&0.145 &0.145    \\\hline
    3 & 1 & 0.348 & 0.246& 0.249& 0.341&0.341 &0.341    \\
    \hline \hline
    \end{tabular}
}
\begin{tablenotes}
      \small
      \item LOCUS+LR: a two-step composite pipeline that first uses LOCUS to extract latent connectivity traits from observed brain connectivity data using without utilizing the clinical information, then predicts the clinical outcome using subjects' trait loadings derived from LOCUS using linear regression.
    \end{tablenotes}
\label{tab:data-driven-sim}
\end{table}

\begin{table}[h!] 
\caption{Parameter estimation root mean square error (RMSE) in small-scale and data-driven simulations. }
\subfloat[Small-scale simulation training root mean square error (RMSE) of the estimation of $\mathbf{U}$, $\boldsymbol{\Lambda}$, $\mathbf{d}$, $\sigma^2$, and $\tau^2$ under varying population sizes, network and clinical signal-to-noise ratios.\label{tab:par-estimation-1}]{
\begin{tabular}{lll|lllll}
    \hline \hline
    \multicolumn{3}{c}{Simulation setting} &\multicolumn{5}{c}{RMSE} \\ \cline{1-3} \cline{4-8}
    $M$ & $SNR_Y$ & $SNR_C$ & $\mathbf{U}$ & $\boldsymbol{\lambda}$ &$\sigma^2$ & $\mathbf{d}$ & $\tau^2$ \\    \hline
    300 & 0.5 & 3 & 2.976 & 0.416 & 1.058 & 0.437 & 1.018 \\ \hline
    500 & 0.5 & 3 & 2.975 & 0.400 & 1.057 & 0.305  & 0.981\\ \hline 
    1000 & 0.5 & 3 & 3.633 & 0.406 & 1.070 & 0.492 & 1.113\\ \hline 
    500 & 0.1 & 6 & 3.620 & 0.589 & 0.908 & 0.158 & 0.073\\ \hline 
    500 & 0.5 & 6 & 2.975 & 0.400 & 1.057 & 0.152 & 0.066 \\
    \hline \hline
    \end{tabular}
}

\subfloat[Data-driven simulation training root mean square error (RMSE) of the estimation of $\mathbf{U}$, $\boldsymbol{\Lambda}$, $\mathbf{d}$, and $\tau^2$ under varying network and clinical signal-to-noise ratios.\label{tab:par-estimation-2}]{
\begin{tabular}{ll|llll}
    \hline \hline
    \multicolumn{2}{l}{Simulation setting} & \multicolumn{4}{c}{RMSE $\times$ 100}\\ \cline{1-2} \cline{3-6}
    $SNR_Y$ & $SNR_C$ &  $\mathbf{U}$ &$\boldsymbol{\lambda}$ & $\mathbf{d}$ & $\tau^2$ \\    \hline
     0.0461 & 3  &  206.964 
&1.603 & 33.138 & 26.665 \\ \hline
     0.5 & 3  &  206.967 
&1.047 & 33.313 & 26.791 \\ \hline 
     1 & 3  &  253.481 
&1.014 & 33.360 & 26.807\\ \hline 
     3 & 3  &  253.481 
&0.990 & 33.374 & 26.836  \\ \hline 
     3 & 0.38  & 253.481 
&0.990 & 116.465 &29.280   \\ \hline
     3 & 0.5 & 253.481 
&0.990 & 101.459 &28.656 \\ \hline 
     3 & 1 & 253.481 &0.990 & 67.164 &27.653 \\
    \hline \hline
    \end{tabular}
}

\end{table}

\section{Application to PReDICT Study}
We applied BSNMani to the clinical and brain connectivity data from the predictors of remission in depression to individual and combined treatments (PReDICT) study. The PReDICT study (\cite{dunlop2012predictors,dunlop2018differential}) is conducted by the Emory University Mood and Anxiety Disorders Program (MAP). The study enrolled adults (18-65 years) with major depressive disorder (MDD) and examined the efficacy of antidepressant medications in improving depressive symptoms. For 12 weeks, patients were randomized to received antidepressant medications (escitalopram or duloxetine) or cognitive behavioral therapy (CBT). The severity of the depressive symptoms of the patients was measured using HDRS (Hamilton Depression Rating Scale \cite{hamilton1960rating}). Additional clinical data such as age, gender, race, education level, etc. were also collected. In addition to clinical data, we also have the patients' rs-fMRI measurements. The Schaefer brain atlas (\cite{schaefer2018local}) was used to parcellate the neuroimaging data into 400 ROIs in the cortical area, with an additional 54 ROIs on the subcortical area from the Melbourne Subcortex Atlas \cite{tian2020topographic}. The functional brain connectivity data was processed via Fisher's Z transformation prior to computational analysis. There are 130 patients for whom brain connectivity data and clinical information are available. Patients' rs-fMRI measurements and clinical HDRS scores were measured before treatment and 12 weeks after treatment. For our analysis, we analyze the baseline z-transformed brain connectivity data for the patients, and examined the association between connectivity traits and the difference in severity of MDD (measured by HRSD) between baseline and week 12 measurements, while adjusting for additional clinical covariates: age, gender, treatment type (anti-depressive medication or CBT), and response group (response or no response to treatment). 

We assessed BSNMani's predictive performance through ten repeated 5-fold cross validation. Under different random seeds, we split the dataset into 5 folds. For training, we applied BSNMani to the baseline network and clinical data and updated all model parameters. For testing, we update the posterior samples of the subject-specific subnetwork summary features loadings $\{\boldsymbol{\lambda}_i\}_{i=1}^M$, using the training samples of the remaining model parameters. Posterior predictive samples of the clinical outcome were then generated for each testing subject. We applied similar training and testing procedures to LOCUS+LR, lasso, and elastic net (SVR and GroupBoosting were excluded from comparison due to their extensive runtime at the PReDICT study's brain network dimensions). The predictive performance of all models is evaluated by comparing their corresponding predictive $R^2$.

\begin{figure}[h!]
\centering
\includegraphics[scale=0.45]{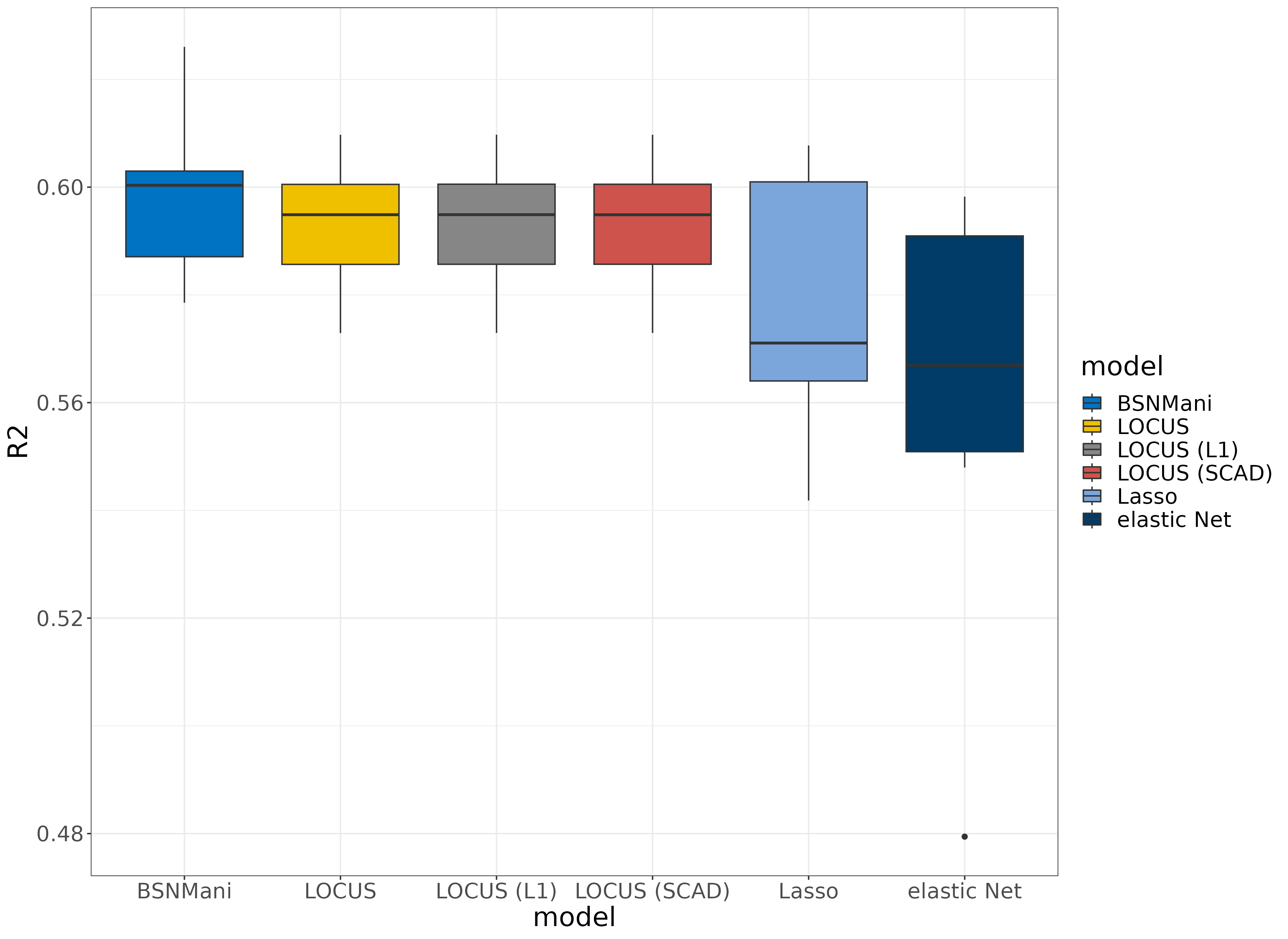}
\caption{Predictive performance comparison over repeated 5-fold cross validation of BSNMani ($q=4$) with existing methods: LOCUS+LR* under default SCAD regularization, L1 regularization, and no regularization, respectively, lasso, and elastic net.} 
\begin{tablenotes}
      \small
      \item LOCUS+LR: a two-step composite pipeline that first uses LOCUS to extract latent connectivity traits from observed brain connectivity data using without utilizing the clinical information, then predicts the clinical outcome using subjects' trait loadings derived from LOCUS using linear regression.
    \end{tablenotes}
\label{5CV_RMSE}
\end{figure}

Our results indicate that BSNMani achieved superior predictive performance compared to all the additional methods that we tested. As shown in Figure \ref{5CV_RMSE}(a), BSNMani achieved the highest overall predictive accuracy among all models, with a median predictive R2 of around 0.6. This is followed by the composite LOCUS+LR approaches, whose median predictive R2 is slightly below 0.6. The full edge set regression methods such as lasso and elastic net tended to perform worse than methods such as BSNMani and LOCUS+LR that can recover subnetworks. Lasso and elastic net achieved a lower median predictive R2 of just above 0.56. We also observed that the subnetwork methods had overall more stable performance than the full edge set regression methods.

\begin{figure}[h!]
\centering
\includegraphics[scale=0.5]{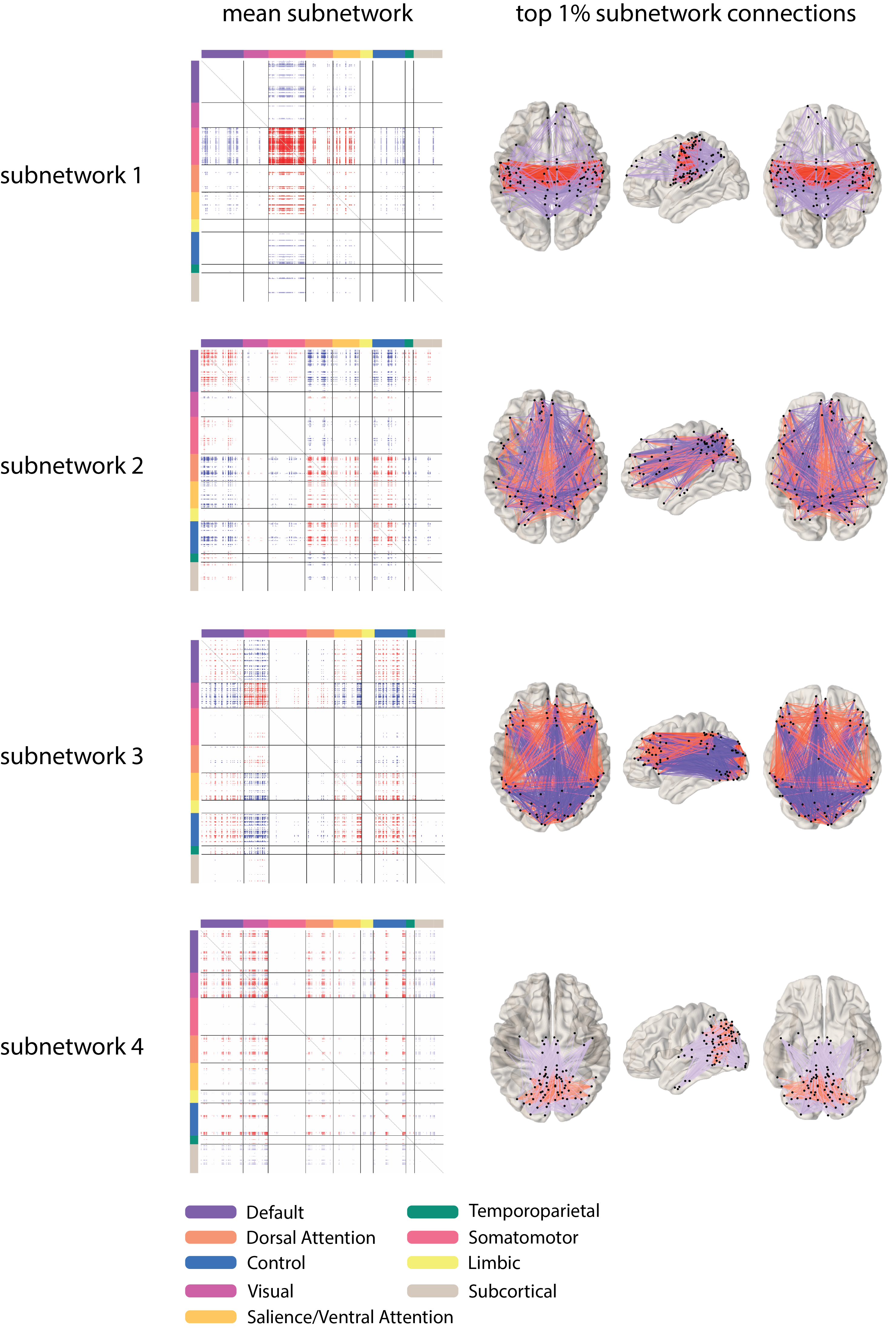}
\caption{Recovered average subnetworks for $q=4$. Each subnetwork is visualized from left to right via a heatmap of the mean subnetwork (thresholded at top 5\%) and the top 1\% subnetwork connections visualized over the top, left and bottom view of the brain (blue edges represent negative connections, red edges represent positive connections).}
\label{PReDICT_subnet}
\end{figure}

\begin{figure}[h!]
    \centering
    \includegraphics[scale=0.25]{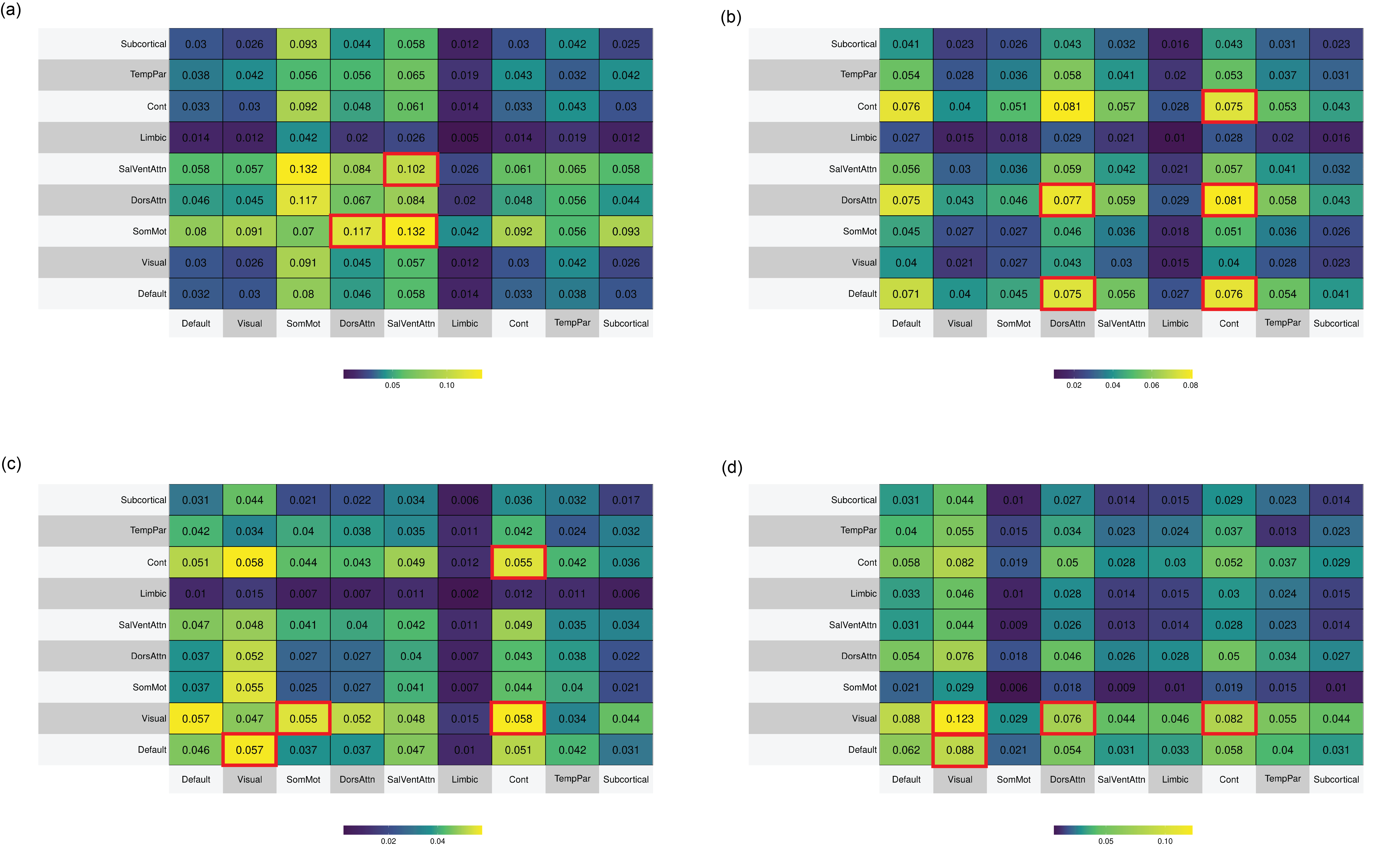}
    \caption{The effect of one SD (standard deviation) increase in connectivity between functional module pairs on observed functional connectivity.}
    \label{fig:subnet_effects}
\end{figure}

We further demonstrate the advantage of BSNMani over full edge set regression approaches by examining the recovered subnetworks corresponding to $q=4$. Figure \ref{PReDICT_subnet} and \ref{fig:subnet_effects} show that BSNMani recovered biologically meaningful underlying subnetworks that reveal various functional connectivity mechanisms related to changes in the severity of MDD. Figure \ref{PReDICT_subnet} shows each mean subnetwork, as well as the top 1\% of the subnetwork connections throughout the brain. The effect of connectivity between functional module pairs on the observed functional connectivity is shown in Figure \ref{fig:subnet_effects}. Subnetwork 1 mainly consists of connections within the somatomotor module and between the somatomotor and the dorsal and salience/ventral attention modules. Subnetwork 2 consists mainly of connections within and between the default module and the dorsal attention, salience/ventral attention, and control modules. Subnetwork 3 recovers a dense hub-like structure around the visual module, as well as connectivity to the salience/ventral attention, and the control and temporoparietal modules. Finally, subnetwork 4 shows a dense hub between and within the default and the visual modules. The subnetworks recovered by BSNMani reveal interesting insights into the underlying connectivity mechanism underlying patients' response to MDD treatment. Our recovered subnetworks match those in separate studies on MMD remission (\cite{dunlop2023shared}, \cite{mac2023individuals}). Specifically, the recovered subnetwork connectivity within and between the Default, Control, and Salience/Ventral Attention networks play an important role in MDD remission, as shown by \cite{dunlop2023shared}'s study on changes in functional brain connectivity after MDD remission. The specific effect of the subject network summary features $\boldsymbol{\lambda}$ on the change in HRSD score is shown in Table \ref{tab:lambda_effect_size}, where the summary features of subnetwork 1 has positive effect on MDD remission and subnetworks 2 to 4 have negative effects on MDD remission. For example, 1 SD (standard deviation) increase in $\lambda_1$ lead to 0.053 decrease in change in HRSD score, and 1 SD increase in $\lambda_2$ lead to 0.684 increase in change in HRSD). 

\newpage

\newpage
\begin{table}[h!] 
\caption{The effect of one SD (standard deviation) increase in $\boldsymbol{\lambda}$ on the change in HRSD score between baseline and week 12.}
    \begin{tabular}{l|llll}
    \hline \hline
    & $\lambda_1$ & $\lambda_2$ & $\lambda_3$ & $\lambda_4$\\\hline
    $\Delta_C$ & -0.053 & 0.684 & 0.094 & 0.520 \\\hline\hline
    \end{tabular} 
\label{tab:lambda_effect_size}
\end{table}

\newpage

\newpage
\section{Discussion}
This work proposes a novel Bayesian scalar-on-network regression model motivated by the increasing importance of brain connectivity networks in the study of mental illness. BSNMani adopts a novel joint model approach, using both brain connectivity and clinical information to infer latent subnetworks that underlie functional connectivity in the brain. Such a framework inherently captures both population-wide connectivity patterns and subject-specific loadings with respect to the clinical outcome of interest. Our low-rank approximation approach to inferring the brain network structure allows for more efficient computation, which is advantageous compared to existing strategies that directly use the full network edge set to model clinical outcomes, where large, dense edge sets lead to less stable results. BSNMani inherently links the population-wide subnetwork structures with the predictors in the clinical prediction model for clear interpretation, avoiding any issues of label-switching. This provides additional advantage to blind source separation methods for brain connectivity such as LOCUS and connICA which are not designed for clinical prediction. BSNMani's novel two-stage hybrid posterior sampling alogrithm efficiently obtains posterior samples of model parameters from a very high-dimensional joint posterior distribution. Our algorithm only require the gradient of the target distribution and avoids costly computation of the Hessian. This poses a distinct advantage compared to Riemann manifold based approaches (\cite{girolami2011riemann},\cite{welling2011bayesian}), which are commonly used to sample from high dimensional distributions. Through real data application to the PReDICT study, we demonstrated BSNMani's advantage in predictive performance with regards to predicting clinical outcome, as well as its advantage in simultaneously recovering meaningful underlying subnetworks related to clinical phenotypes using relatively few latent dimensions. We further validated BSNMani's robustness through thorough simulation studies, where we showed BSNMani is capable of accurately estimating model parameters and maintaining high predictive accuracy under high levels of noise in the data and model mis-specification. 




\begin{supplement}
\stitle{Supplementary A.} \sdescription{The supplementary material document contains details on BSNMani's joint posterior sampling algorithm, and the two-stage posterior sampling algorithm.}
\end{supplement}


\bibliographystyle{imsart-nameyear} 
\bibliography{references}       
\end{document}


\begin{frontmatter}
\title{SUPPLEMENTARY MATERIALS TO THE PAPER: "BSNMani: Bayesian Scalar-on-network Regression with Manifold Learning"}

\begin{aug}
\author[A]{\fnms{Yijun}~\snm{Li}\ead[label=e1]{liyijun@umich.edu}\orcid{0000-0003-0513-9565}},
\author[B]{\fnms{Ki Sueng}~\snm{Choi}\ead[label=e2]{kisueng.choi@mssm.edu}\orcid{0000-0002-6446-0677}},
\author[C]{\fnms{Boadie W.}~\snm{Dunlop}\ead[label=e3]{
bdunlop@emory.edu}\orcid{0000-0002-4653-0483}},
\author[C]{\fnms{W. Edward}~\snm{Craighead}\ead[label=e8]{
ecraigh@emory.edu}\orcid{0000-0001-9957-0027}},
\author[B]{\fnms{Helen S.}~\snm{Mayberg}\ead[label=e4]{
helen.mayberg@mssm.edu}\orcid{0000-0002-1672-2716}},
\author[E]{\fnms{Lana}~\snm{Garmire}\ead[label=e7]{lgarmire@med.umich.edu}\orcid{0000-0002-4654-2126}},
\author[D]{\fnms{Ying}~\snm{Guo}\ead[label=e5]{yguo2@emory.edu}\orcid{0000-0003-3934-3097}},
\author[A]{\fnms{Jian}~\snm{Kang}\ead[label=e6]{jiankang@umich.edu}\orcid{0000-0002-5643-2668}}
\address[A]{Department of Biostatistics,
University of Michigan\printead[presep={,\ }]{e1};\printead{e6}}
\address[B]{Icahn School of Medicine at Mount Sinai, New York, NY USA\printead[presep={,\ }]{e2};\printead{e4}}
\address[C]{Department of Psychiatry and Behavioral Sciences, Emory University School of Medicine\printead[presep={,\ }]{e3}\printead{e8}}
\address[E]{Department of Computational Medicine and Bioinformatics,
University of Michigan \printead[presep={,\ }]{e7}}
\address[D]{Department of Biostatistics and Bioinformatics,
Emory University\printead[presep={,\ }]{e5}}
\end{aug}

\end{frontmatter}



\section{Derivation of $\nabla \mathrm{log}\pi(\mathbf{X}|\sim)$ for Joint Posterior Sampling}
Define the following functions: \begin{align*}
    R(\mathbf{X}) &= (\mathbf{X}^T\mathbf{X})^{-1/2}\\
    A_i(\mathbf{U}) &= A_i(\mathbf{U}_{\mathbf{X}}) = \boldsymbol{\Lambda}_i\mathbf{U}^T_{\mathbf{X}}\mathbf{Y}_i\\
    F_i(\mathbf{U}) &= F_i(\mathbf{U}_{\mathbf{X}}) = \mathbf{A}_i\mathbf{U}_{\mathbf{X}}
\end{align*}
Define $\otimes$ as the Kronecker product, and $\boxtimes$ as the box product, such that given a $m_1\times n_1$ matrix $\mathbf{G}$ and a $m_2 \times n_2$ matrix $\mathbf{H}$, $\mathbf{G}\boxtimes\mathbf{H}=\begin{bmatrix}
    g_{11}h_{11}& g_{12}h_{11} &\cdots & g_{1n_1}h_{1n_2}\\
    \vdots & \vdots & \ddots &\vdots\\
    g_{11}h_{m_21} & g_{12}h_{m_21} & \cdots & g_{1n_1}h_{m_2n_2}\\
    \vdots & \vdots & \ddots &\vdots \\
    g_{m_11}h_{m_21}& g_{m_12}h_{m_21} & \cdots & g_{m_1n_1}h_{m_2n_2}
\end{bmatrix}$.\\
We can then derive $\nabla \mathrm{log}\pi(\mathbf{X}|\sim)$ as \begin{align*}
    \nabla\mathrm{log}&\pi(\mathbf{X}|\sim) = -\mathbf{X}+\frac{\partial}{\partial\mathbf{X}}\mathrm{log}\pi_{\mathbf{U}}(\mathbf{U}_{\mathbf{X}})\\
    \frac{\partial}{\partial\mathbf{X}}&\mathrm{log}\pi_{\mathbf{U}}(\mathbf{U}_{\mathbf{X}}) = \frac{\partial}{\partial\mathbf{X}}\mathrm{trace}(F(\mathbf{U})) = \mathrm{vec}^{-1}(\mathrm{vec}^T(\mathbf{I}_q)(\sigma^{-2}\sum_{i=1}^M\frac{\partial}{\partial\mathbf{U}}F_i(\mathbf{U})\cdot\frac{\partial}{\partial\mathbf{X}}\mathbf{U}))\\
    \frac{\partial}{\partial\mathbf{U}}&F_i(\mathbf{U})= (\mathbf{I}_q\otimes\mathbf{U}^T)(\boldsymbol{\Lambda}_i\boxtimes\mathbf{Y}_i)+((\boldsymbol{\Lambda}_i\mathbf{U}^T_{\mathbf{X}}\mathbf{Y}_i)\otimes\mathbf{I}_q)\mathbf{I}_{Nq}\\
    \frac{\partial}{\partial\mathbf{X}}&\mathbf{U}= (\mathbf{I}_n\otimes(\mathbf{X}^T\mathbf{X})^{-1/2})\mathbf{I}_{Nq}+(\mathbf{X}\otimes\mathbf{I}_q)\frac{\partial}{\partial\mathbf{X}}\mathbf{R}\\
    \frac{\partial}{\partial\mathbf{X}}&\mathbf{R} = -\left(\mathbf{I}_q\otimes(\mathbf{X}^T\mathbf{X})^{-1/2}+(\mathbf{X}^T\mathbf{X})^{-1/2}\otimes\mathbf{I}_q\right)^{-1}\left((\mathbf{X}^T\mathbf{X})^{-1}\otimes(\mathbf{X}^T\mathbf{X})^{-1}\right)\cdot\\
    &\left((\mathbf{I}_q\otimes\mathbf{X}^T)(\mathbf{I}_q\boxtimes\mathbf{I}_N)+(\mathbf{X}^T\otimes\mathbf{I}_q)\mathbf{I}_{Nq}\right)
\end{align*}
\section{Derivation of $\prod_{i=1}^M\pi(\mathbf{Y}_i|\boldsymbol{\Theta}_{Y})$ for Two-stage Posterior Sampling}
\begin{align*}\pi(\mathbf{Y}_i|\boldsymbol{\Theta}_Y,\tau^2_{\boldsymbol{\lambda}}) &=\int \pi(\mathbf{Y}_i|\boldsymbol{\Theta}_Y,\boldsymbol{\lambda}_i)\pi(\boldsymbol{\lambda}_i|\tau^2_{\boldsymbol{\lambda}})d\boldsymbol{\lambda}_i\\
    &= \int C_{\mathbf{Y}_i}(\sigma^2)^{-\frac{1}{4}N(N+1)}\mathrm{etr}\left[-\frac{1}{2}(\sigma^2)^{-1}(\mathbf{Y}_i-\mathbf{U}\boldsymbol{\Lambda}_i\mathbf{U}^T)(\mathbf{Y}_i-\mathbf{U}\boldsymbol{\Lambda}_i\mathbf{U}^T)\right] \cdot\\
    & C_{\boldsymbol{\lambda}_i}(\tau^2_{\boldsymbol{\lambda}})^{-\frac{q}{2}}\mathrm{exp}\left[-\frac{1}{2}(\tau^2_{\boldsymbol{\lambda}})^{-1}\boldsymbol{\lambda}_i^T\boldsymbol{\lambda}_i\right]d\boldsymbol{\lambda}_i\\
    &=C_{\mathbf{Y}_i}C_{\boldsymbol{\lambda}_i}(\sigma^2)^{-\frac{1}{4}N(N+1)}(\tau^2_{\boldsymbol{\lambda}})^{-\frac{q}{2}}\mathrm{exp}\left[-\frac{1}{2}(\sigma^2)^{-1}||\mathbf{Y}_i||^2_F\right]\cdot\\
    &\mathrm{exp}\left[\frac{1}{2}\boldsymbol{\mu}_0^T\boldsymbol{\Sigma}_0^{-1}\boldsymbol{\mu}_0\right](2\pi)^{\frac{q}{2}}(\sigma^{-2}+\tau^{-2}_{\boldsymbol{\lambda}})^{-\frac{q}{2}},
\end{align*}
where $\boldsymbol{\Sigma}_0 = (\sigma^{-2}+\tau^{-2}_{\boldsymbol{\lambda}})^{-1}$, and $ \boldsymbol{\mu}_0 = \frac{\sigma^{-2}}{\sigma^{-2}+\tau^{-2}_{\boldsymbol{\lambda}}}\mathrm{diag}(\mathbf{U}^T\mathbf{Y}_i\mathbf{U})$. The network data likelihood is shown below:
\begin{align*}
\prod_{i=1}^M\pi(\mathbf{Y}_i|&\boldsymbol{\Theta}_{Y}) = \int \left[\prod_{i=1}^M \pi(\mathbf{Y}_i|\boldsymbol{\Theta}_Y,\tau^2_{\boldsymbol{\lambda}})\right]\pi(\tau^2_{\boldsymbol{\lambda}})d\tau^2_{\boldsymbol{\lambda}}\\
&=C^M_{\mathbf{Y}_i}C^M_{\boldsymbol{\lambda}_i}C_{\tau^2_{\boldsymbol{\lambda}}}(2\pi)^{\frac{qM}{2}}(\sigma^2)^{-\frac{1}{4}MN(N+1)}\mathrm{exp}\left[-\frac{1}{2}(\sigma^2)^{-1}\sum_{i=1}^M||\mathbf{Y}_i||_F^2\right]\cdot\\
& \int (\tau^2_{\boldsymbol{\lambda}})^{-\frac{qM+\eta_0}{2}-1}(\sigma^{-2}+\tau^{-2}_{\boldsymbol{\lambda}})^{-\frac{qM}{2}}\cdot\\
&\mathrm{exp}\left[\frac{1}{2}\frac{(\sigma^{-2})^{2}}{\sigma^{-2}+\tau^{-2}_{\boldsymbol{\lambda}}}\left(\sum_{i=1}^M\mathrm{diag}^T(\mathbf{U}^T\mathbf{Y}_i\mathbf{U})\mathrm{diag}(\mathbf{U}^T\mathbf{Y}_i\mathbf{U})\right)-\frac{1}{2}\eta_0\tau^2_0(\tau^2_{\boldsymbol{\lambda}})^{-1}\right]d\tau^2_{\boldsymbol{\lambda}}.
\end{align*}
\subsection{Numerical Approximation of $\prod_{i=1}^M\pi(\mathbf{Y}_i|\boldsymbol{\Theta}_{Y})$ with Sparse Grid Integration}
 Since $\prod_{i=1}^M\pi(\mathbf{Y}_i|\boldsymbol{\Theta}_{Y})$ has no closed form, we can use Sparse Grid Integration to approximate it by decomposing the integral into the following general form
$$\prod_{i=1}^M\pi(\mathbf{Y}_i|\boldsymbol{\Theta}_{Y}) = \int h(x)w(x) dx,$$
We can decompose the integral into a weighted sum with Gaussian weights. We make the following change of variable: \begin{align*}
    x&=\mathrm{log}(\tau^2_{\boldsymbol{\lambda}}), \\
    w(x)&=\frac{1}{\sqrt{2\pi}}\mathrm{exp}\left(-\frac{x^2}{2}\right), \\
    h(x)&=\sqrt{2\pi}\cdot \mathrm{exp}\left[\left(-\frac{qM+\eta_0}{2}+2\right)x+\frac{x^2}{2}\right](\sigma^{-2}+\mathrm{exp}(-x))^{-\frac{qM}{2}}\cdot\\
    &\mathrm{exp}\left[\frac{1}{2}\frac{(\sigma^{-2})^2}{\sigma^{-2}+\mathrm{exp}(-x)}\left(\sum_{i=1}^M\mathrm{diag}^T(\mathbf{U}^T\mathbf{Y}_i\mathbf{U})\mathrm{diag}((\mathbf{U}^T\mathbf{Y}_i\mathbf{U}))\right)-\frac{1}{2}\eta_0\tau^2_0\mathrm{exp}(-x)\right].
\end{align*} 
Given a specified accuracy level $k$, Sparse Grid will pre-compute the $R$ grid values $x$ and the corresponding weights $w$. The integral can then be computed as $\sum_{r=1}^R h(x_r)w_r$. The overall likelihood can be written as \begin{align*}
    \prod_{i=1}^M\pi(\mathbf{Y}_i|\boldsymbol{\Theta}_{Y}) &= C^M_{\mathbf{Y}_i}C^M_{\boldsymbol{\lambda}_i}C_{\tau^2_{\boldsymbol{\lambda}}}(2\pi)^{\frac{qM}{2}}(\sigma^2)^{-\frac{1}{4}MN(N+1)}\cdot\\
    &\mathrm{exp}\left[-\frac{1}{2}(\sigma^2)^{-1}\sum_{i=1}^M||\mathbf{Y}_i||_F^2\right]\cdot \left(\sum_{r=1}^R h(x_r)w_r\right),
\end{align*}
where $x_r$ and $w_r$ are the Sparse Grid nodes and weights.

\section{Derivation of $A(\boldsymbol{\lambda})$}
The updated normalizing constant for stage-two is shown below:
\begin{align*}
    A(\boldsymbol{\lambda}) = \int\left[\prod_{i=1}^M\pi(C_i|\boldsymbol{\Theta}_{C},\boldsymbol{\lambda}_i)\right]\pi(\boldsymbol{\Theta}_{C}|\tau^2_{\boldsymbol{\alpha}},\tau^2_{\boldsymbol{\beta}})\pi(\tau^2_{\boldsymbol{\alpha}})\pi(\tau^2_{\boldsymbol{\beta}})d\boldsymbol{\Theta}_{C}d\tau^2_{\boldsymbol{\alpha}}d\tau^2_{\boldsymbol{\beta}}
\end{align*}

Denote $\mathbf{d}=[\boldsymbol{\beta}^T, \boldsymbol{\alpha}^T]^T$, $\mathbf{x}_i = [\boldsymbol{\lambda}^T_i,\mathbf{z}^T_i]^T$, and $\mathbf{W} = \begin{bmatrix}
\tau^2_{\boldsymbol{\beta}}\mathbf{I}_q & \mathbf{0}\\
\mathbf{0} & \tau^2_{\boldsymbol{\alpha}}\mathbf{I}_r
\end{bmatrix}$. We integrate out $\mathbf{d}$ and simplify $\boldsymbol{A}(\boldsymbol{\lambda})$: \begin{align*}
    A(\boldsymbol{\lambda}) &= (2\pi)^{\frac{q+r}{2}}C_c^MC_{\boldsymbol{\alpha}}C_{\boldsymbol{\beta}}C_{\tau^2}C_{\tau^2_{\boldsymbol{\alpha}}}C_{\tau^2_{\boldsymbol{\beta}}}\int\int\int(\tau^2)^{-1-\frac{M+\rho_0}{2}}(\tau^2_{\boldsymbol{\alpha}})^{-1-\frac{r+\omega_0}{2}}(\tau^2_{\boldsymbol{\beta}})^{-1-\frac{q+\gamma_0}{2}}\cdot\\
    &\mathrm{exp}\left[-\frac{1}{2}\left(\rho_0\psi^2_0(\tau^2)^{-1}+\omega_0\phi^2_0(\tau^2_{\boldsymbol{\alpha}})^{-1}+\gamma_0\kappa^2_0(\tau^2_{\boldsymbol{\beta}})^{-1}+(\tau^2)^{-1}\left(\sum_{i=1}^MC_i^2\right)\right)\right]\cdot\\
    &\mathrm{exp}\left[\frac{1}{2}(\tau^2)^{-2}\left(\sum_{i=1}^MC_i\mathbf{x}^T_i\right)\left[\mathbf{W}^{-1}+(\tau^2)^{-1}\left(\sum_{i=1}^M\mathbf{x}_i\mathbf{x}_i^T\right)\right]^{-1}\left(\sum_{i=1}^MC_i\mathbf{x}_i\right)\right] \cdot \\
    &\mathrm{det}\left(\mathbf{W}^{-1}+(\tau^2)^{-1}\left(\sum_{i=1}^M\mathbf{x}_i\mathbf{x}_i^T\right)\right)^{-\frac{1}{2}}d\tau^2d\tau^2_{\boldsymbol{\alpha}}d\tau^2_{\boldsymbol{\beta}}.
\end{align*}

\subsection{Approximation of $A(\boldsymbol{\lambda})$ with Sparse Grid Integration}
The closed-form for $A(\boldsymbol{\lambda})$ is very cumbersome. We can approximate the above integral using Sparse Grid Integration by decomposing it into the following general form:
$$\int\int\int h(x_1,x_2,x_3;\boldsymbol{\lambda})w(x_1)w(x_2)w(x_3)dx_1dx_2dx_3.$$
We can decompose the integral as an unweighted sum of $h(x_1,x_2,x_3)$ over the pre-computed grid. Since the integration needs to be over $[0,1]^3$, we do the following change of variables: \begin{align*}
    x_1&=\mathrm{exp}(-\tau^{-2}), \qquad x_2 = \mathrm{exp}(-\tau^{-2}_{\boldsymbol{\alpha}}), \qquad x_3 = \mathrm{exp}(-\tau^{-2}_{\boldsymbol{\beta}}),\\
    w(x_1)& = 1, \qquad w(x_2)=1, \qquad w(x_3)=1,
\end{align*}
\begin{align*}
    &\resizebox{\textwidth}{!}{h(x_1,x_2,x_3)=(\mathrm{log}(x_1^{-1}))^{\frac{M+\rho_0}{2}-1}(\mathrm{log}(x_2^{-1}))^{\frac{r+\omega_0}{2}-1}(\mathrm{log}(x_3^{-1}))^{\frac{q+\gamma_0}{2}-1}x_1^{\left[\frac{1}{2}\left(\rho_0\psi^2_0+\sum_{i=1}^MC_i^2\right)-1\right]}x_2^{\frac{1}{2}\omega_0\phi^2_0-1}x_3^{\frac{1}{2}\gamma_0\kappa^2_0-1}}\cdot\\
    &\resizebox{\textwidth}{!}{\mathrm{exp}\left[\frac{1}{2}(\mathrm{log}(x_1^{-1}))^2\left(\sum_{i=1}^MC_i\mathbf{x}_i^T\right)\left[\begin{bmatrix}
        \mathrm{log}(x_3^{-1})\mathbf{I}_q & \mathbf{0}\\
        \mathbf{0} & \mathrm{log}(x_2^{-1})\mathbf{I}_r \end{bmatrix}+\mathrm{log}(x_1^{-1})\left(\sum_{i=1}^M\mathbf{x}_i\mathbf{x}_i^T\right)\right]^{-1}\left(\sum_{i=1}^MC_i\mathbf{x}_i\right)\right]}\cdot\\
        &\mathrm{det}\left(\begin{bmatrix}
        \mathrm{log}(x_3^{-1})\mathbf{I}_q & \mathbf{0}\\
        \mathbf{0} & \mathrm{log}(x_2^{-1})\mathbf{I}_r \end{bmatrix}+\mathrm{log}(x_1^{-1})\left(\sum_{i=1}^M\mathbf{x}_i\mathbf{x}_i^T\right)\right)^{-\frac{1}{2}}.
\end{align*}
Sparse Grid will generate a set of $R$ nodes and corresponding weights for each dimension. $A(\boldsymbol{\lambda})$ can then be approximated as $$\tilde{A}(\boldsymbol{\lambda}) = (2\pi)^{\frac{q+r}{2}}C_c^MC_{\boldsymbol{\alpha}}C_{\boldsymbol{\beta}}C_{\tau^2}C_{\tau^2_{\boldsymbol{\alpha}}}C_{\tau^2_{\boldsymbol{\beta}}}\cdot\sum_{r=1}^Rw(x_{r1})w(x_{r2})w(x_{r3})h(x_{r1},x_{r2},x_{r3}),$$
where $x_{r1}$, $x_{r2}$, and $x_{r3}$ are the sparse grid nodes in each dimension. 

$\mathrm{sup}_{\{\boldsymbol{\lambda}_i\}}A(\boldsymbol{\lambda})$ can be approximated via numerical optimization techniques such as the Broyden–Fletcher–Goldfarb–Shanno algorithm (BFGS).